# A robust generalizable device-agnostic deep learning model for sleep-wake determination from triaxial wrist accelerometry

Nasim Montazeri[1,2b], Stone Yang[2,3a], Dominik Luszczynski[2,3a], John Zhang[,2,3a], Dharmendra Gurve[2], Andrew Centen[2], Maged Goubran[2a,c,3b], Andrew Lim[2b,c,3c]

[1]Queen's University Dept. of Electrical & Computer Engineering Kingston, ON, Canada
[2]Sunnybrook Research Institute, [a]Physical Sciences, [b]Dept. of Medicine Neurology Div., [c]Hurvitz Brain Sciences, Toronto, ON, Canada
[3]University of Toronto Dept. of [a] Computer Science, [b]Medical Biophysics, [c]Medicine, Toronto, ON, Canada

Corresponding author: Andrew S. P. Lim
andrew.lim@utoronto.ca

**ABSTRACT**

**Study Objectives:** Wrist accelerometry is widely used for inferring sleep-wake state. Previous works demonstrated poor wake detection, without cross-device generalizability and validation in different age range and sleep disorders. We developed a robust deep learning model for to detect sleep-wakefulness from triaxial accelerometry and evaluated its validity across three devices and in a large adult population spanning a wide range of ages with and without sleep disorders.

**Methods:** We collected wrist accelerometry simultaneous to polysomnography (PSG) in 453 adults undergoing clinical sleep testing at a tertiary care sleep laboratory, using three devices. We extracted features in 30-second epochs and trained a 3-class model to detect wake, sleep, and sleep with arousals, which was then collapsed into wake vs. sleep using a decision tree. To enhance wake detection, the model was specifically trained on randomly selected subjects with low sleep efficiency and/or high arousal index from one device recording and then tested on the remaining recordings.

**Results:** The model showed high performance with F1 Score of 0.86, sensitivity (sleep) of 0.87, and specificity (wakefulness) of 0.78, and significant and moderate correlation to PSG in predicting total sleep time (R=0.69) and sleep efficiency (R=0.63). Model performance was robust to the presence of sleep disorders, including sleep apnea and periodic limb movements in sleep, and was consistent across all three models of accelerometer.

**Conclusions:** We present a deep model to detect sleep-wakefulness from actigraphy in adults with relative robustness to the presence of sleep disorders and generalizability across diverse commonly used wrist accelerometers.

**The Statement of Significance:**

We developed a deep learning model that can robustly detect sleep-wake state using wrist triaxial accelerometry across three different devices in adults of diverse ages, with and without sleep disorders.

**Introduction**

Wrist accelerometry is well tolerated and capable of recording continuously for weeks at a time in natural settings[1]. As there is an approximate correspondence between lack of movement and sleep, actigraphy has been extensively used for long-term monitoring of rest/activity patterns and sleep/wake state.

Making accurate inferences about sleep/wake state from actigraphy is not without difficulty. Current algorithms for extracting sleep measures from actigraphy have been validated mainly on omni-directional accelerometers.[2] Furthermore, many current algorithms are heuristic and rely on constant thresholds applied to actigraphy-based indices.[3-8] These algorithms have been mostly developed and validated on healthy young adults, rather than adults with sleep disorders or older adults with medical co-morbidities[4,5,9-14], which can lead to differences in model performance.[5,15] Furthermore, generalizability across different actigraphy devices is often not well established[16], and existing models have tended to have relatively low sensitivity for wake.

Several AI-based models have emerged to try and address some of these limitations.[4,9,13,17-19] However, these have generally been validated in relatively small samples of healthy and younger adults, and low or unreported wake detection (specificity) remains a challenge.

The primary objective of this study was to develop an AI model to predict sleep/wake state from triaxial actigraphy recordings and validate it against sleep-wake state determined by simultaneous polysomnography (PSG) in a wide range of adults of diverse ages, with and without sleep disorders. The second objective was to demonstrate the generalizability and robustness of the developed AI model across different models of wrist accelerometers.

**Material and methods**

*Participants:* Adults 40 years of age or above, referred to the clinical sleep laboratory at Sunnybrook Health Sciences Centre in Toronto, Canada, for clinical PSG, were recruited for this study. The study protocol was approved by the Research Ethics Board of Sunnybrook Hospital, and all participants provided written consent before participating in the study.

*Data Acquisition:* Wrist accelerometer recordings were obtained from 453 adults undergoing concurrent clinical PSG, which were recorded according to AASM guidelines and with a Grael PSG system (Compumedics, Victoria, Australia). Wrist accelerometers (AX3, Axivity, UK recorded at 25 Hz, n=138; GENEActiv Activinsights, UK recorded at 100 Hz n=224; GT9X, Actigraph, United States, recorded at 100 Hz n=91). Participants were first instrumented with the wrist accelerometers, whose recordings were started ahead of the set-up and initiation of PSG recordings.

Each PSG recording was visually annotated according to AASM guidelines by one of six board-certified polysomnography technologists overseen by a board-certified sleep physician (AL). We extracted sleep/wake state, presence of absence of arousals, and summary measures including the apnea hypopnea index and periodic limb movement index. In addition, the portion of the accelerometer recordings preceding the onset of PSG recording was labeled as wake based on direct observation of participants by study staff.

Raw data from the Axivity and GENEActiv accelerometers were downloaded using the GGIR package in R. Raw data from the GT9X accelerometers were exported into csv format using the Actilife package and then read into R. Accelerometry data were aligned to PSG data by direct visualization of data on the CrowdEEG platform[20] and manual alignment of arousals.

*Data Analysis:*

**Preprocessing:** Signals for each axis ($G_x$, $G_y$, $G_z$) were smoothed using a median filter with 5 second length and 2.5 second overlap. Then, the angle between z component and xy plane [$Angle_z$= atan ($G_z$/ $(G_x^2 + G_y^2)^{0.5}$)], the angle between x component and yz plane [$Angle_x$= atan ($G_x$/ $(G_y^2 + G_z^2)^{0.5}$)] and Euclidean Norm Minus One [ENMO = $(G_x^2 + G_y^2 + G_z^2)^{0.5}$ -1] were calculated. We noticed that the three accelerometers used here differed in their sensitivity to acceleration values near zero. To render these more comparable, we subtracted from all ENMO values an ad hoc threshold of 0.029g and any resulting negative values were set to zero. Then, ENMO extracted from the three devices was bandpass filtered using a three-rank Butterworth filter with a bandwidth between 0.5 and 3 Hz. The raw temperature was lowpass filtered using a two-rank Butterworth filter with a cutoff of 0.005 Hz. Finally, the raw fluctuations in each dimension were extracted by removing the baseline of each raw accelerometer signal. The baseline changes based on the body position, which generates distinguishable steps in the accelerometry signal. These steps were detected as baseline change points using a widely used technique based on linear optimization and used to segment each signal.[21] This algorithm considers several potential changepoints and optimize their number and locations based on the pruned exact linear time (PELT) method.[22] For each segment, 50th percentile was considered as baseline value and subtracted from the signal within that segment. Then the segments were sequentially concatenated to obtain an aligned version of the accelerometer signal with removed baseline ($A_x$, $A_y$, $A_z$).

**Feature extraction:** For each 30-second epoch, the corresponding segment of actigraphy data was isolated, from which four types of features were extracted:

- Activity features ($f_1$-$f_4$, where $f_i$ denotes $i^{th}$ feature): The average value of smoothed temperature relative to the 95$^{th}$ percentile for that recording, ENMO, Angle$_x$ and Angle$_z$ were extracted.
- Morphological features ($f_5$-$f_{31}$): Each thirty-second epoch was divided into first, middle and last ten-second intervals. Then, from each interval, three previously proposed features including zero-crossing rate (ZCR) relative to the 95$^{th}$ percentile for the recording, time-above-threshold (TAT), and proportional-integrating mode (PIM) were extracted from $A_x$, $A_y$ and $A_z$.[23]
- Frequency features ($f_{32}$-$f_{37}$): Fundamental frequency and its spectral power were calculated for $A_x$, $A_y$ and $A_z$. Fundamental frequency ($FF_i$, i=[x,y,z]) is the lowest frequency in the spectrum of a signal with the highest power.
- Body motion powers ($f_{38}$-$f_{41}$): The power of four different distinguished body motions, including motion related to the heartbeat potentially sensed over the wrist (0.6-2 Hz), slow hand motion (2-3.5 Hz), quick hand motion (3.5-10 Hz) and jerk in the body (10-15 Hz) were extracted.

Overall, forty-one features were extracted from every thirty-epoch of accelerometry data. Moreover, to improve the generalizability of the model across devices, certain features were transformed individually, using scikit-learn. Specifically, ZRC was standardized, while Angle-related features were fitted to their distribution extracted from the entire recordings in the training set for the relevant device. The PIM-related features were transformed by removing the median and scaling them according to the quantile range, while the frequency and body motion features were scaled using a Yeo-Johnson transform, where both scalars were fitted to the training set. For each data epoch, two reference annotations were extracted to mark: 1-sleep-wake and 2- presence-absence of arousals. The extracted actigraphy-based features along with PSG-based annotations were used to train a hierarchical model to detect sleep-wakefulness.

**Hierarchical model sleep detection model:** The hierarchy consisted of two mathematical models:

1. A three-class deep learning (DL) model: A supervised DL model was developed to use the extracted features of each thirty-epoch and classify it to one of the three classes: sleep without arousal, sleep with arousal and wakefulness. This model output was a probability for each class, indicating the likelihood of the epoch's resemblance to the data of that class. The architecture of the deep model consisted of four convolutional layers in a 1-dimensional convolutional neural network (CNN). These layers are known to learn the patterns of the features.[24] To keep the temporal order of the input signal, we selected causal (dilated) padding. In each convolutional layer, the input data were convolved with a kernel of size 4x64 and a stride of one. Each convolutional layer was followed by batch normalization, ReLu activation function, and dropout with a probability of 0.2 to prevent overfitting. Next, the convolutional layers were followed by one layer of Long Short-Term Memory (LSTM).[25] This layer kept track of information flow over time. The LSTM considers the relationship between input and output, to choose whether to remember or forget any flowing information. We included 128 hidden units for every time step with Tanh activation, followed by additive Gaussian noise, ReLu activation function and dropout with a probability of 0.2. The values of the hidden units of LSTM's final outputs were fed to a dense (fully connected) layer with sigmoid activation function to calculate the probability of three classes: sleep, arousal and wakefulness.
2. A two-class decision tree: A decision tree was designed to receive the generated probabilities by the three-class deep model and classify them to sleep or wakefulness. The rationale for developing a hierarchy rather than a single model was related to the fact that an arousal in an epoch may be associated with a sudden body motion. Thus, arousal can increase the likelihood of the epoch being misclassified as wakefulness, even though most arousals are short (<3 seconds) and mostly do not result in a change in sleep state.

This model was implemented using the Keras Python deep learning library version 2.2.4 with TensorFlow GPU backend version 1.12.0 and CUDA 9.1. Our model was trained on ubuntu 16.04 with Intel Core i7 processor with one GTX 1080 GPU with 8GB of GRAM.

**Model Training:** During training, an Adam optimizer[26] was used. The learning rate and weight decay were set to 0.001 and 0.0001, respectively. The batch size was set to 128. Initialization was performed using the Xavier uniform initializer.[27] To specifically improve performance on wake detection, the model was trained specifically on Axivity recordings with total sleep time (TST) <6 hours and apnea hypopnea index (AHI) >0 and with either sleep efficiency <70% or arousal index (arousals per hour) in NREM sleep (NTOAI)>20. A four-fold cross-validation strategy was applied in which three folds were randomly selected to be used as training set and the remaining fold was combined with the remaining subjects with TST >6 hours or AHI = 0 or had sleep efficiency >70% and NTOAI <20 to construct the Axivity test set. From the training set, we randomly selected 15% of the subjects and used them as the validation set. In each training iteration, the two hierarchical models were trained on the remaining training set and validated on the validation set, and the internal parameters were updated. Kernel regularization, dropout, and an early stopping mechanism were applied to reduce overfitting. We implemented early stopping whereby training was stopped if the validation loss did not decrease for eighty consecutive epochs.

**Accelerometry-Only Model:** Not all contemporary wrist accelerometers have a temperature sensor. Therefore, we repeated the above, training and testing a model using only the accelerometry signals.

**Statistical analysis:** Statistical analyses were implemented in R (4.3.2) software. The test sets' demographics, except for sex, were compared to the training set using the t-test or the Wilcoxon rank-sum test, depending on the normality. The Shapiro test was used to check normality. The $\chi 2$ test was applied to compare the number of females versus males in test sets compared to the training set.

The performance of the model with all the features was quantified using evaluation metrics, including F1 score, sensitivity (sleep detection), specificity (wakefulness detection), accuracy, precision, and Cohen's kappa. The metrics were computed both for each test set as a whole and separately for each recording in each test set. The same metrics were extracted from a well-known algorithm called the GGIR[8] for comparison, using a time threshold of 5 minutes and an angle threshold of 5 minutes. We compared the distribution of recording-level metrics between the current model and GGIR using t-tests. To examine the impact of clinical and demographic factors on model performance, we took Cohen's kappa for each individual recording in each test set and used linear regression to consider this as a function of age (<=65 vs <65), sex (male vs. female), presence of sleep apnea (AHI>=15 vs. <15), and presence of periodic limb movements in sleep (PLMI>=5 vs. <5). For comparisons, changes in the selected metric, standard error (SE) and p-value were reported.

Ground truth total sleep time (TST) and sleep efficiency were extracted from the PSG annotations, and predicted values were obtained from the accelerometry data and compared using correlation analysis and Bland-Altman plots. The fragmentation index, defined as the number of sleep intervals less than 5 minutes divided by the total sleep time, was extracted from the detected scores, and compared to the similar index of PSG. Moreover, the length of the detected sleep intervals was extracted from the GGIR algorithm and our proposed method, and the statistical distribution of these intervals was compared to PSG using the Kolmogorov-Smirnov test.

Feature changes were analyzed by calculating their average value during four states: sleep without arousals, sleep with arousals, mobile wakefulness, and motionless wakefulness. The Analysis of Variance (ANOVA) test was used to find any differences, followed by a post hoc analysis to compare their values during sleep without arousals compared to the three other groups.

**Results**

*Characteristics of participants.* Table 1 shows the characteristics of the study participants in the Axivity, GENEActiv, and GT9X test sets. Across all three test sets, the mean [range] age was 57.21 [40.64 - 73.78], 48% were female, 79 [21%] had moderate-severe sleep apnea (defined as AHI>=15), 113 [29%] had significant periodic limb movements in sleep (defined by PLMS>=5).

*Performance evaluation.*
Figure 2 shows an example trace of a subset of the actigraphy-based features and the predicted sleep-wakefulness scores in comparison to the PSG-based hypnogram.

Among the 621 hours of Axivity recordings in the test set, the model had an F1 score of 0.90, sensitivity of 0.91, specificity of 0.79, and Cohen's kappa of 0.71, showing similar sensitivity to GGIR (-0.01), but with considerably improved specificity (+0.11) and kappa (+0.09). Similar differentials in global performance (specificity, F1, and kappa) compared to GGIR were seen in the GENEActiv and GT9x test sets (Table II).

We then computed performance measures for each recording individually and compared the distribution of recording-level performance metrics between the current model and the GGIR algorithm. Across n=381 individual recordings from 3 different accelerometers, overall mean F1, accuracy, precision, Cohen's kappa, and specificity were higher for our DL model vs. GGIR (F1 difference: +0.01, SE=0.01, $p=0.11$; accuracy difference: +0.03, SE=0.01, $p<0.001$, precision difference: +0.06, SE=0.01, $p<0.001$;

kappa difference: +0.07, SE=0.01, p<0.001; specificity difference: +0.13, SE=0.01, p<0.001). However, sensitivity was slightly lower (difference: -0.05, SE=0.01, p<0.001).

*Distribution of Individual Features and Feature Importance*

By analyzing the feature importance in differentiating among sleep without arousals, sleep with arousals, mobile wakefulness, and motionless wakefulness, we found a significant difference in fundamental frequencies of the three actigraphy dimensions ($f_{32-34}$) and power of the heart-related activity ($f_{38}$) during motionless wakefulness compared to sleep intervals, consistently in all three datasets. Whereas previous common features such as PIMs, ZCR and TAT were not consistent in showing significant differences (see supplementary document).

*Effect of Clinical and Demographic Variables*

To examine the effect of clinical and demographic factors on model performance, we used linear regression model to consider recording-level Cohen's kappa as a function of age (>= 65 vs. <65), sex (male vs. female), presence of sleep apnea (AHI>=15 vs. <15) and presence of PLMS (PLMI>=5 vs. <5) for each of the three accelerometers. Model performance, as indicated by Cohen's kappa, was robust to the presence of PLMS and sleep apnea in all three accelerometers (p>0.05 for all). Model performance was robust to age in the GENEActiv and GT9X test sets (p>0.05 for both) but modestly lower in adults over the age of 65 the Axivity set (Kappa difference=-0.13, SE=0.05, p=0.01). Model performance was similar in males vs. females in the Axivity and GT9X test sets (p>0.05 for both) but slightly lower in males vs. females in the GENEActiv test set (Kappa difference =-0.06, SE=0.02, p=0.008).

*Sleep summary measures.*

Aggregating data from all three test sets, there were moderate to strong correlations between PSG- and model-inferred total sleep time (R= +0.69, p<0.001), sleep efficiency (R=+0.63, p<0.001) and sleep fragmentation index (R= +0.46, p<0.001; figure 4).

*Statistical distribution of sleep intervals*

We next computed the distribution of runs of sleep as measured by PSG or accelerometry with either the current model or GGIR. Whereas the distribution of sleep bout lengths detected by the current model was statistically indistinguishable from that detected by PSG (p>0.20 for each of the three devices), the distribution inferred by GGIR differed significantly from PSG for both the Axivity (p=0.005) and GT9x (p=0.001) and there was a non-significant difference for GENEActiv (p=0.06).

*Accelerometer-Only Model*

To examine the importance of different sensors to model performance, we next considered a model using only the accelerometer data, without temperature features. We computed recording-level performance statistics and compared their distribution between our full model (including temperature data) and the reduced model (without temperature data). Overall performance was marginally worse in the model without temperature data (mean F1 0.85 with vs. 0.84 without temperature, SE=0.01, p=0.155; Mean Cohen's kappa 0.66 with temperature and 0.60 without temperature; SE=0.01, p<0.001). Sensitivity was largely preserved (Mean sensitivity 0.87 with temperature and 0.90 without, SE=0.01, p<0.001), but specificity dropped somewhat (mean specificity 0.78 with temperature, 0.69 without temperature, SE=0.01, p<0.001) while remaining higher than that seen with GGIR (mean specificity 0.69 vs. 0.65 for GGIR, SE=0.01, p=0.002).

**Discussion**

In this study, we used a large dataset of 453 concurrent wrist accelerometry and PSG recordings from older adults undergoing concurrent clinical PSG, to develop and validate a DL model to predict sleep-wake state from triaxial wrist accelerometry (with or without skin temperature). Key attributes of the model include: 1) improved specificity (wake detection) compared to many other models, 2) accurate representation of the distribution of sleep bout lengths, 3) validation in a sleep clinic population spanning a wide range of ages and sleep co-morbidities, 4) robustness to the presence of sleep disorders, and 5) generalizability across multiple models of commonly used wrist triaxial accelerometers.

*Overall Performance*

As outlined in Table V, a common challenge with existing models for sleep-wake prediction from wrist accelerometry is poor wake detection (i.e. specificity). Accurate detection of wake is of particular interest in clinical populations as increased wakefulness during the sleep period is a common consequence of sleep disorders as varied as insomnia, sleep apnea, and restless legs; can be associated with neurodegenerative disorders,[28-30] and can be a predictor of adverse clinical outcomes, including cognitive impairment and dementia.[31-33] The current model achieves over 78% specificity across a wide range of adults with clinical sleep complaints, and this is accompanied by relatively high F1 and kappa as well. The relatively high specificity may have been due in part to our having trained the model specifically on a subset of the Axivity data with a higher proportion of wake and arousals – specifically a subset with TST < 6 hours, sleep efficiency <70%, and arousal index >20. As well, we included features such as fundamental frequencies of the three actigraphy dimensions ($f_{32-34}$) and power of the heart related activity ($f_{38}$), which showed significant differences during motionless wakefulness compared to sleep intervals, consistently in all three datasets.

Related to the above, there is also increasing interest in the statistical distribution of sleep bout length durations as a marker of rest/sleep fragmentation[34] which has been associated with numerous adverse neurological outcomes.[34-40] In contrast to another commonly used algorithm, the current model predicts a distribution of bout length durations statistically indistinguishable from concurrent PSG.

Accurate detection of wakefulness ensures the validity of sleep architecture data — a feature often missing in methods based solely on body motion. The relatively high specificity (wakefulness detection) of the current model, with accompanying relatively high kappa and F1, compares favorably to many other models (Table IV). One exception is the model of Ode and colleagues with a reported specificity of 80%, sensitivity of 96%, and F1 of 89% as compared to PSG.[41] However, we note that the model from Ode and colleagues was trained and tested only on a small (n=20) set of young healthy volunteers without psychiatric, sleep, or neurological disorders. Moreover, we note that some of the participants in this study contributed more than one recording to the dataset, which in the context of cross-validation used in this study, may have resulted in overestimation of model performance. It remains uncertain whether this model would perform similarly well in older adults with sleep complaints who typically present to clinical sleep disorder centers.

*Generalizability*

Indeed, one of the strengths of the present study is the broad demographic and clinical diversity of the study population (mean age 57, 49%>age 60; 22% with AHI>15; 30% with PLMI>5), and the large size (n=381) of the test set, which stands in contrast to many models that are developed and tested exclusively on younger adults without sleep complaints[4,9-13]. Of note, model performance was robust to the presence/absence of sleep apnea or periodic limb movements in sleep, and robust to differences in age and sex when applied to two of the three accelerometers.

Another dimension of generalizability is the relatively consistent performance when applied to different models of accelerometer. We tested the current model on data from 3 different commonly used models of wrist accelerometer, with similar performance across devices. This contrasts with many existing models which were developed and tested on a single model of accelerometer leaving their generalizability uncertain.

*Acceleration Only Model*

Previous studies have reported that the addition of temperature data can substantially improve the specificity of sleep-wake inference.[42] We observed a similar effect here – a model trained without temperature features had similar overall F1 (0.84 vs. 0.85) and preserved sensitivity (0.90 vs. 0.87) but with a drop in specificity (0.69 vs. 0.78). This may be because the rise in skin temperature that accompanies sleep may be one feature the model uses to distinguish resting wakefulness from sleep.

*Limitations*

A few methodological limitations are worth considering. First, the recordings were performed overnight, with a brief period of wakefulness before lights out. As such, there was a preponderance of sleep vs. wake. It is unclear how well the model would perform at detecting daytime naps, for instance, during periods dominated by wake. Second, all recordings were performed at a single site, and generalizability to other settings remains to be confirmed. Third, all recordings were obtained from adults presenting to a clinical sleep disorders centre. Additional work is needed to confirm generalizability to younger, healthier individuals, although they are presumably easier to classify due to the lack of neurodegenerative disorders, co-morbidities, and lower incidence of sleep disturbances.

*Conclusions*

This study introduces a CNN/LSTM model to automatically predict sleep/wake state from wrist triaxial accelerometry with or without skin temperature data. In a large test set of individuals presenting to a tertiary care clinical sleep laboratory spanning a wide range of ages and sleep-related co-morbidities, and applied to three different models of accelerometer, the model displayed high overall performance. Notably, the model demonstrated improved specificity compared to other models in the literature, good reproduction of the statistical distribution of sleep bout lengths, robustness to sleep disorders like sleep apnea and periodic limb movements in sleep, applicability to clinical populations, and generalizability across multiple commonly used wrist accelerometers. The presented model provides an important improvement in the ability of wrist accelerometry to provide accurate wake/sleep inference in clinical settings.

**Data Availability statement:** Data used in this study will be made available upon request.

**Financial Disclosure Statement:** This study was supported by grants from the Centre for Aging Brain Health Innovation, National Institute on Aging, and Canadian Institutes of Health Research. The authors have no financial conflicts of interest to disclose.

**Non-financial Disclosure Statement:** The authors have no non-financial conflicts of interest to disclose.

Table I –Demographic characteristics of study participants

| Characteristics | Training set (Axivity) | Test set (Axivity) | Test set (GENEActiv) | Test set (GT9X) |
|---|---|---|---|---|
| Sample size (# female) | 72 (20) | 66 (30) | 224 (112) | 91 (28) |
| Age (yr) | 59 ± 15 | 51 ± 17** | 58 ± 16 | 60 ± 16 |
| Apnea-hypopnea index (AHI, events/hr) | 6.55 [1.0 – 33.2] | 1.4 [0.1 – 6.1]*** | 4.9 [1.7 – 14.5]* | 7.0 [1.2 – 14.3]* |
| Body mass index (BMI, kg/m$^2$) | 30.38 ± 7.7 | 29.53 ± 6.4 | 30.42 ± 7.47 | 28.1 ± 5.6* |
| Periodic Limb Movement Syndrome (events, PLMS) | 2 [0 - 6] | 1 [0 - 5]* | 0 [0 - 4] | 2 [0 - 19]*** |
| Sleep efficiency (%) | 66 ± 15 | 80 ± 11*** | 69 ± 19 | 70 ± 16* |

The results are presented in mean ± standard deviation for characteristics with normal distribution and in median [1st quartile – 3rd quartile]. * p-value ≤0.05, ** p-value≤0.01, *** p-value ≤0.001.

Table II: Global performance of the developed model over different test sets compared to the GGIR algorithm.

| | Characteristics | F1-score (%) | Sensitivity (%) | Specificity (%) | Precision (%) | Accuracy (%) | Cohen's Kappa | Recording Duration (Hr) | Number of Recordings |
|---|---|---|---|---|---|---|---|---|---|
| Current model | Axivity | 89.98 | 91.20 | 79.26 | 88.79 | 86.94 | 0.712 | 621.58 | 66 |
| | GENEActiv | 83.65 | 86.55 | 78.09 | 80.94 | 82.47 | 0.648 | 2107.22 | 224 |
| | GT9X | 86.34 | 86.30 | 77.05 | 86.38 | 82.85 | 0.633 | 746.84 | 91 |
| | Combined | **85.56** | 87.44 | **78.07** | **83.76** | **83.35** | **0.659** | 3475.64 | 381 |
| GGIR | Axivity | 87.90 | 92.29 | 68.15 | 83.91 | 83.66 | 0.629 | 621.58 | 66 |
| | GENEActiv | 82.17 | 92.10 | 65.52 | 74.17 | 79.29 | 0.581 | 2107.22 | 224 |
| | GT9X | 85.39 | 90.93 | 62.83 | 80.49 | 80.47 | 0.563 | 746.84 | 91 |
| | Combined | 83.96 | **92.01** | 65.09 | 77.21 | 80.23 | 0.587 | 3475.64 | 381 |

Table III – Distribution of individual-level performance metrics of the developed model over different test sets compared to the GGIR algorithm.

| | Characteristics | F1-score (%) (SD) | Sensitivity (%) (SD) | Specificity (%) (SD) | Precision (%) (SD) | Accuracy (%) (SD) | Cohen's Kappa (SD) | Recording Duration (Hr) (SD) | Number of Recordings |
|---|---|---|---|---|---|---|---|---|---|
| Current model | Axixity | 89.69 (5.57) | 91.56 (6.84) | 79.47 (13.54) | 88.68 (8.69) | 86.99 (6.24) | 0.712 (0.146) | 9.42 (0.80) | 66 |
| | GENEActiv | 81.68 (13.07) | 86.40 (12.16) | 79.46 (14.74) | 80.38 (16.95) | 82.51 (9.43) | 0.635 (0.179) | 9.41 (0.64) | 224 |
| | GT9X | 84.34 (13.64) | 85.45 (14.18) | 77.37 (16.51) | 85.30 (15.01) | 82.75 (8.7) | 0.606 (0.195) | 8.21 (0.62) | 91 |
| | **Combined** | **83.70 (12.59)** | 87.07 (12.12) | **78.96 (14.97)** | **82.99 (15.68)** | **83.34 (8.92)** | **0.639 (0.180)** | 9.12 (0.84) | 381 |
| GGIR | Axixity | 87.58 (6.06) | 92.79 (8.18) | 67.82 (15.50) | 83.90 (8.89) | 83.73 (6.91) | 0.615 (0.170) | 9.42 (0.80) | 66 |
| | GENEActiv | 80.29 (13.30) | 91.90 (10.21) | 66.59 (15.98) | 73.62 (16.81) | 79.36 (10.54) | 0.571 (0.184) | 9.41 (0.64) | 224 |
| | GT9X | 83.44 (13.58) | 90.24 (13.31) | 62.46 (19.25) | 79.74 (15.26) | 80.35 (10.81) | 0.536 (0.217) | 8.21 (0.62) | 91 |
| | Combined | 82.30 (12.70) | **91.66 (10.74)** | 65.82 (16.81) | 76.86 (15.85) | 80.35 (10.18) | 0.570 (0.191) | 9.12 (0.62) | 381 |

For each algorithm and test set, the average ± standard deviation of metrics and the related global metrics are reported in the first and second rows, respectively.

Table IV: Kappa score differences (ΔK) between groups defined for age, apnea-hypopnea index (AHI), Periodic limb movement syndrome (PLMS) and sex on test sets of Axivity, GENEActiv and GT9X. standard error (SE) and p-value (p) reported for each comparison.

| Category | Axivity | | | Geneactiv | | | GT9X | | |
|---|---|---|---|---|---|---|---|---|---|
| | ΔK | SE | p | ΔK | SE | p | ΔK | SE | p |
| Apnea (AHI >= 15) | 0.123 | 0.064 | 0.058 | -0.035 | 0.028 | 0.210 | -0.041 | 0.050 | 0.412 |
| Age (Age >= 65) | -0.126 | 0.047 | 0.010 | -0.034 | 0.019 | 0.165 | -0.047 | 0.044 | 0.289 |
| Limb Movement (PLMS >= 5) | 0.064 | 0.478 | 0.185 | -0.034 | 0.027 | 0.218 | -0.030 | 0.043 | 0.491 |
| Sex | -0.064 | 0.042 | 0.131 | -0.064 | 0.024 | 0.008 | -0.017 | 0.040 | 0.670 |

Table V: Comparison between our work and related literature

| Study | modalities | # subject | method | results |
|---|---|---|---|---|
| M. Mario et al. 2013[5] | Actiwatch AW-64 | 90 patients were considered from the following studies: insomnia; baseline sleep in healthy participants; older adults, daytime sleep in night-workers | Cole-Kripke algorithm | ACC: 86.63%<br>SEN: 96.5%<br>SPC: 32.9%<br>F1: -<br>Precision: -<br>Cohen's kappa: - |
| Kosmadopoulos et al. 2014[4] | Actical Actiwatch-64[18] | 22 young adults (18 male, four females) | ActiLife software | ACC: 83.5%<br>SEN: 87.6%<br>SPC: 61.5<br>F1: -<br>Precision: 89.2%<br>Cohen's kappa: 0.35 |
| Aktaruzzaman et al. 2017[9] | GENEActiv | 18 healthy young subjects | support vector machine (SVM) | ACC: 77%<br>SEN: 82%<br>SPC: 50%<br>F1: -<br>Precision: -<br>Cohen's kappa: 0.3 |
| Cho et al. 2019[17] | GT3X | 10 healthy volunteers, (3 females and 7 males) | CNN + LSTM | ACC: 89.7%<br>SEN: 89.0%<br>SPC: -<br>F1: -<br>Precision: 89.2%<br>Cohen's kappa: 0.60 |
| Haghayegh et al. 2020[18] | Actigraph | 40 healthy adults (17 female, mean age: 26.7 years) | CNN + LSTM | ACC: 87.8%<br>SEN: 94.1%<br>SPC: 64.0%<br>F1: -<br>Precision: -<br>Cohen's kappa: 59.9 |
| Ode et al. 2022[41] | A custom-made accelerometer | 25 healthy adults with repeated trials | XGBoost | ACC: 91.74%<br>SEN: 96.13%<br>SPC: 80.18%<br>F1: 88.87%<br>Precision: -<br>Cohen's kappa: - |
| This work | **Axivity Geneactiv GT9X** | **453 (including people with sleep apnea and PLMS, and older adults)** | CNN + LSTM | ACC: 83.4%<br>SEN: 87.4%<br>**SPC: 78.1%**<br>**F1: 85.6%**<br>Precision: 83.8%<br>**Cohen's kappa: 0.66** |

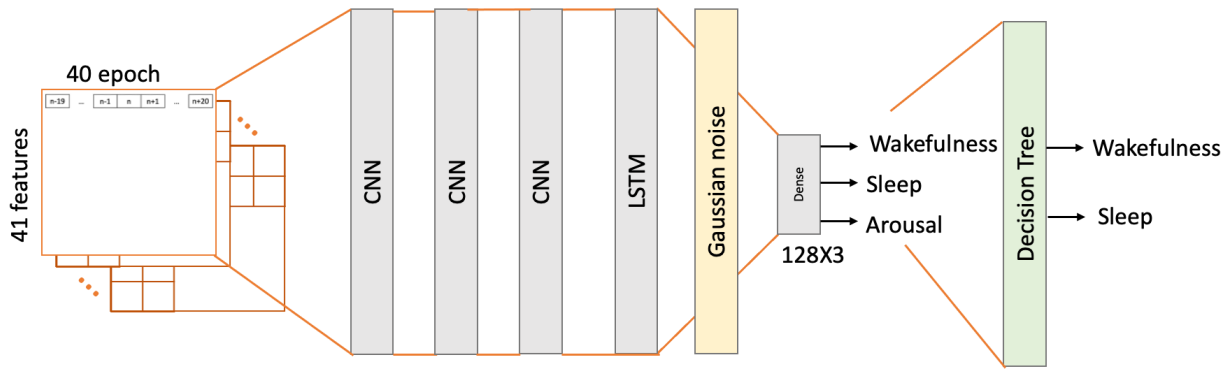

Figure 1: Overview of model architecture and our hierarchical two-model approach. CNN: Convolutional neural network; LSTM: Long short-term memory.

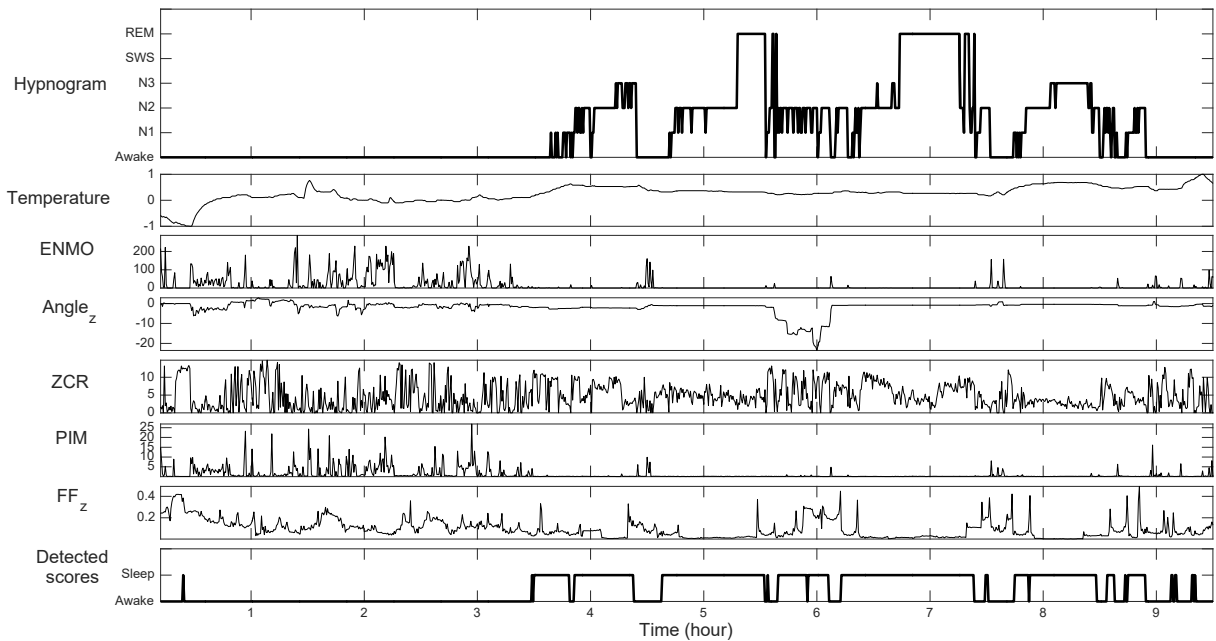

Figure 2: Example trace of a subject with 67% sleep efficiency. The reference polysomnography-based scores are reflected in hypnogram. Five selected features include temperature, Euclidean Norm Minus One (ENMO), the angle between the z-axis and xy plane ($Angle_z$), zero-crossing rate (ZCR), and proportional-integrating mode (PIM), and fundamental frequency of z-axis ($FF_z$). The detected scores by the model with an F1 score of 89.50% are shown in the last panel.

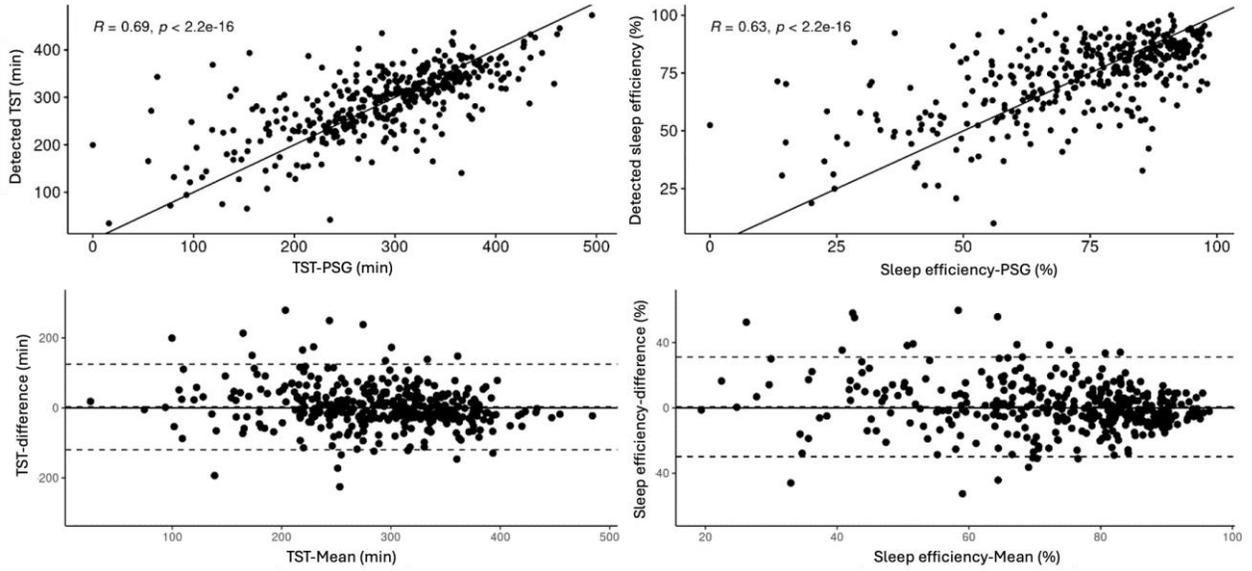

Figure 3: Scatter plots (first row) and Bland-Altman plots (second row) between model-predicted and PSG values for two sleep measures: total sleeping time (TST) and sleep efficiency. Correlation coefficients (R) and p-values (p) are reported for each scatter plot. The points are extracted from all three databases included in our study.

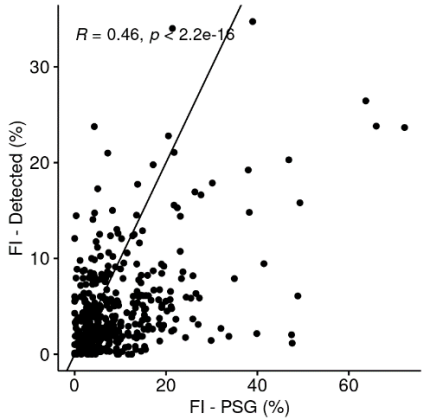

Figure 4: Correlation between model-based and PSG-based Fragmentation Index (FI), defined as the number of sleep intervals less than 5 minutes divided by the total sleep time. The points are extracted from all three databases included in our study.

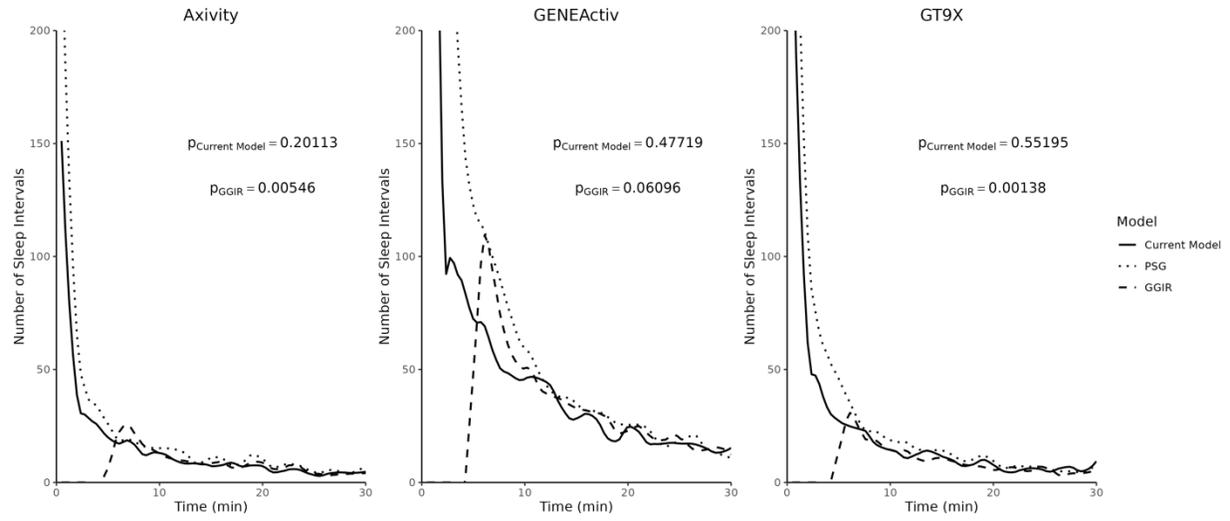

Figure 5: Number of sleep intervals in different lengths detected by the proposed model and GGIR, compared to those extracted from PSG.

# Supplementary document

# A robust generalizable device-agnostic deep learning model for sleep-wake determination from triaxial wrist accelerometry

Nasim Montazeri, Stone Yang, Dominik Luszczynski, John Zhang, Dharmendra Gurve, Andrew Centen, Maged Goubran, Andrew Lim

**Sleep annotations:** In Table S1, we present the results of analyzing agreement among six different sleep technicians in annotating sleep-wakefulness measures. When using the annotations generated by each of the annotators to train the model, we found no significant differences in the F1 score of the test set.

Table S1: Effect of Scoring Technician – Current Model

| Category | Group | Axivity | | Geneactiv | | GT9X | |
|---|---|---|---|---|---|---|---|
| | | F1 Mean (SD) | p-value | F1 Mean (SD) | p-value | F1 Mean (SD) | p-value |
| Technician | TL | 88.65 (7.34) | | 81.85 (11.66) | | N/A | |
| | DJ | 89.61 (5.67) | | 81.97 (13.42) | | 84.66 (12.29) | |
| | LL | 91.96 (4.43) | 0.737 | N/A | 0.545 | N/A | 0.696 |
| | RS | 90.47 (4.09) | | 81.62 (10.02) | | 86.02 (8.92) | |
| | MQB | N/A | | 74.95 (19.86) | | N/A | |
| | MB | N/A | | 84.73 (14.21) | | N/A | |

**The variations of the proposed features:** In Figures S1 to S3, we illustrate the changes in various features and compare them across different states: sleep without arousal, sleep with arousal, mobile wakefulness, and motionless wakefulness. To define wake states, we calculated the 80th percentile of ENMO for each test set. Wake samples with ENMO values above this threshold were labeled as "Mobile Wake," while those below were labeled as "Motionless Wake." These results are discussed in detail in the main text.

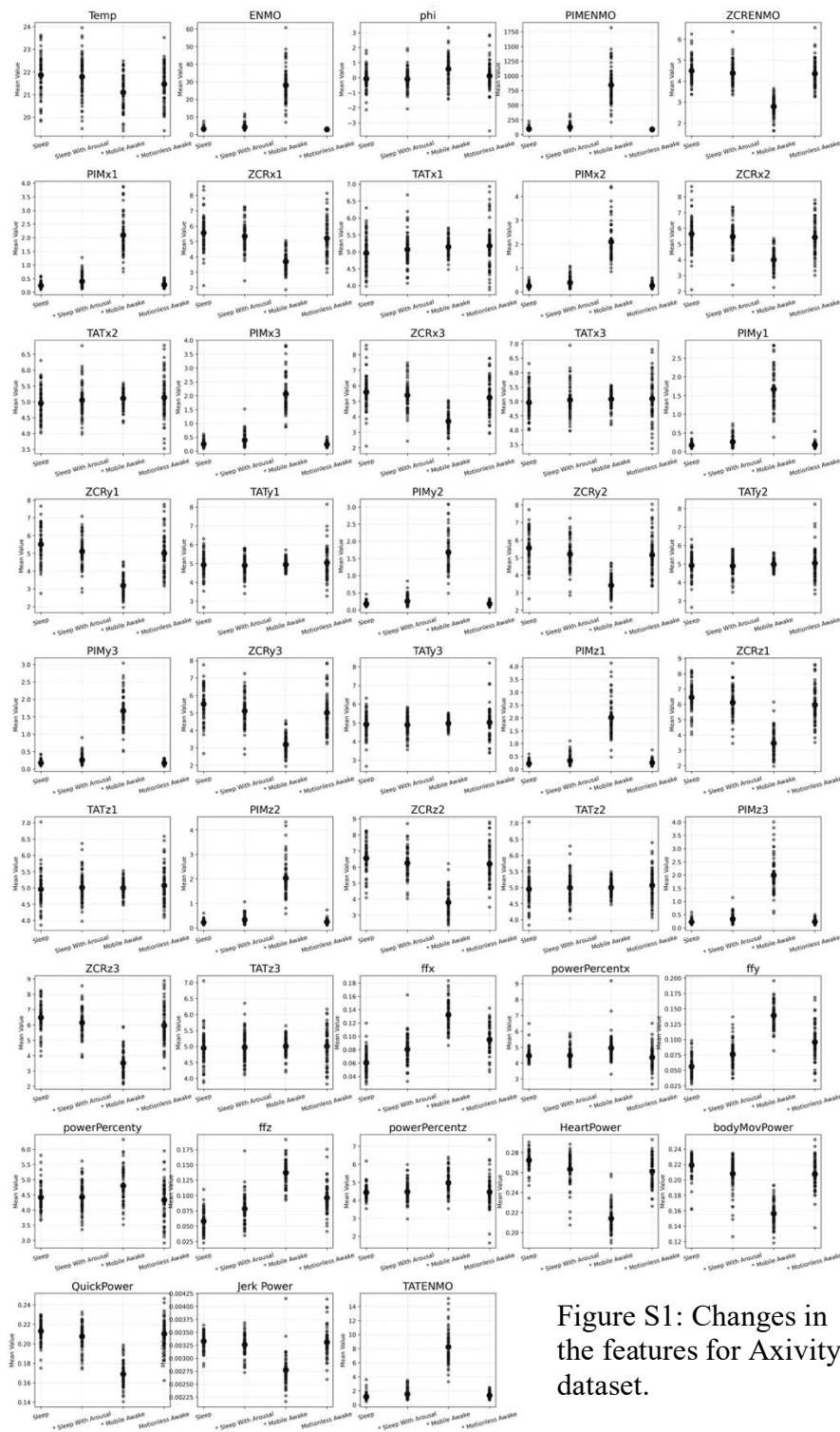

Figure S1: Changes in the features for Axivity dataset.

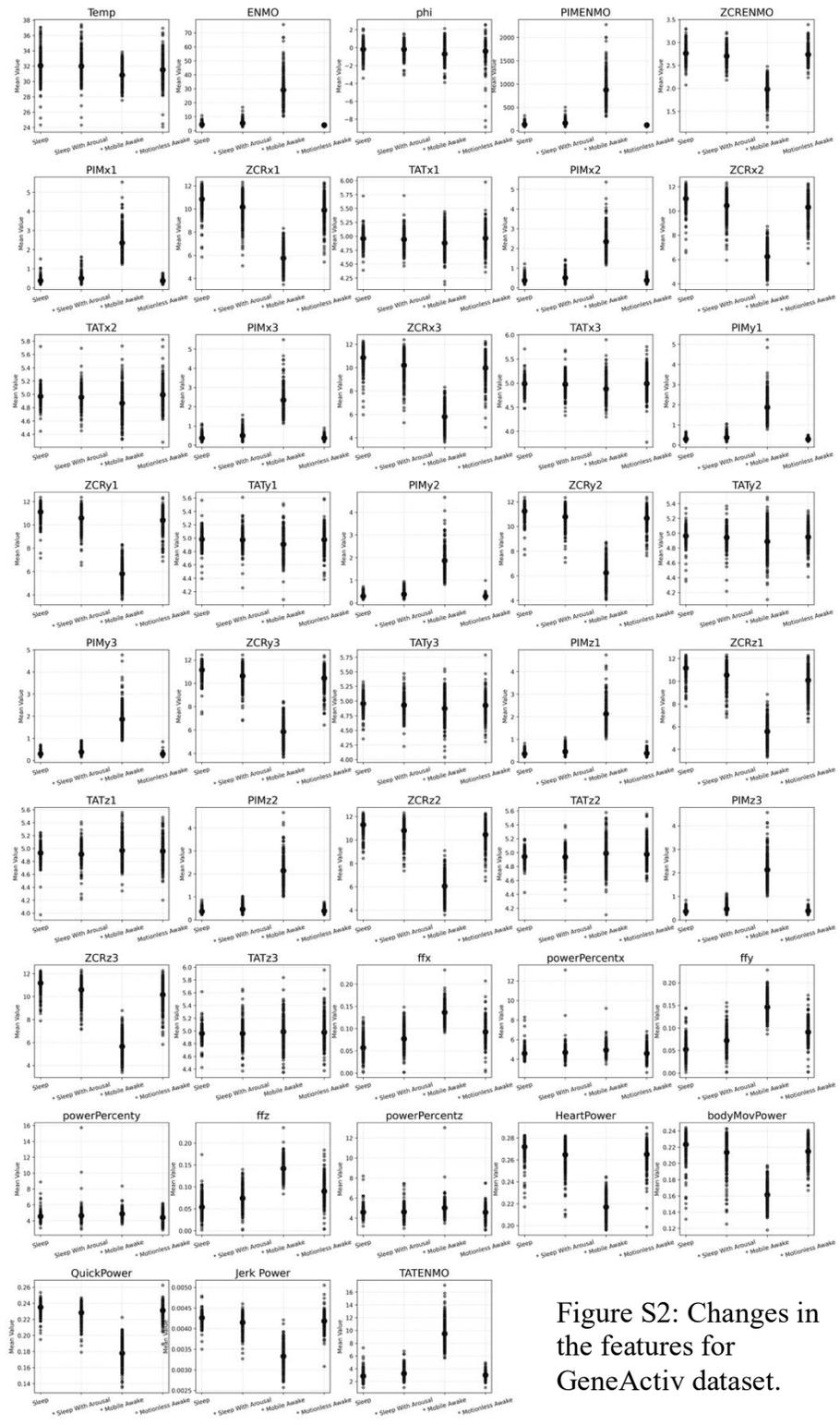

Figure S2: Changes in the features for GeneActiv dataset.

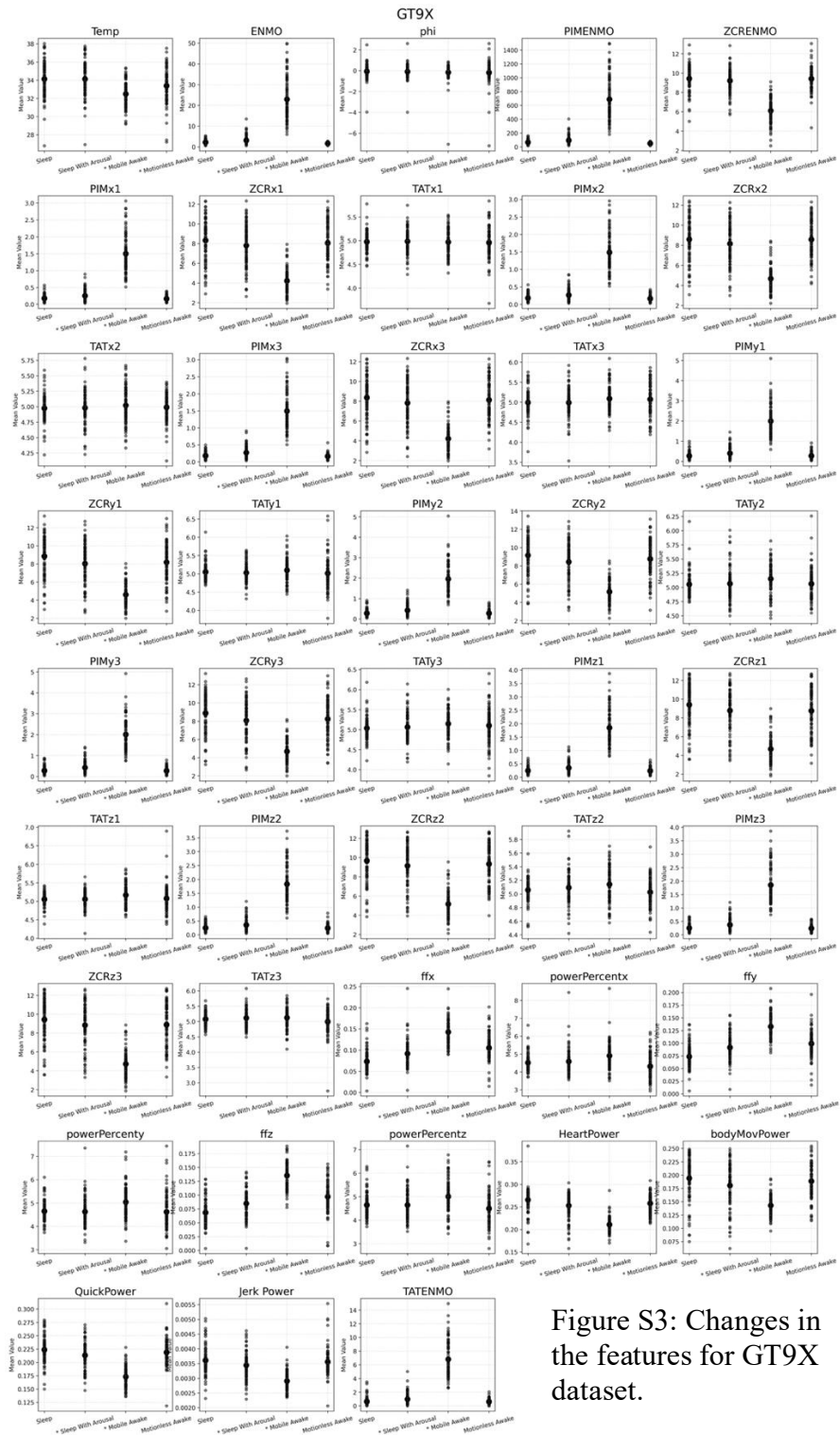

Figure S3: Changes in the features for GT9X dataset.